\documentclass[12pt]{article}

\usepackage{amssymb}
\begin{document}
\baselineskip=18.6pt plus 0.2pt minus 0.1pt

 \def\be{\begin{equation}}
  \def\ee{\end{equation}}
  \def\bea{\begin{eqnarray}}
  \def\eea{\end{eqnarray}}
  \def\nn{\nonumber\\ }
\newcommand{\nc}{\newcommand}
\nc{\al}{\alpha}
\nc{\bib}{\bibitem}
\nc{\la}{\lambda}
\nc{\C}{\mbox{\hspace{1.24mm}\rule{0.2mm}{2.5mm}\hspace{-2.7mm} C}}
\nc{\R}{\mbox{\hspace{.04mm}\rule{0.2mm}{2.8mm}\hspace{-1.5mm} R}}

\begin{titlepage}
\title{
\begin{flushright}
 {\normalsize \small GNPHE/0405}
\\[.1cm]
\mbox{}
\end{flushright}
{\bf Toric Calabi-Yau supermanifolds }\\[.3cm] {\bf and mirror symmetry }}
\author{A. Belhaj$^{1}$\thanks{{\tt abelh633@mathstat.uottawa.ca}} ,\ \
 L.B. Drissi$^{2}$,\ \
 J. Rasmussen$^{3}$\thanks{{\tt rasmusse@crm.umontreal.ca}} ,\ \
 E.H. Saidi$^{2}$\thanks{{\tt esaidi@ictp.trieste.it}} ,\ \ 
 A. Sebbar$^1$\thanks{{\tt sebbar@mathstat.uottawa.ca}}
\\[5pt]
{\it \small $^1$  Department of Mathematics and Statistics,
 Ottawa University}\\{ \it \small 585 King Edward Ave.,
Ottawa, ON, Canada,  K1N 6N5  }\\
{\it \small $^2$ National Grouping of High Energy Physics,
 GNPHE and Lab/UFR  HEP}\\
{\it \small Department of Physics,  Faculty of Sciences, Rabat, Morocco}
\\
{\it\small $^3$ Centre de Recherches Math\'ematiques, Universit\'e de
Montr\'eal} \\
{\it\small C.P. 6128, succursale centre-ville, Montr\'eal, PQ, Canada H3C 3J7}
}
\maketitle \thispagestyle{empty}

\begin{abstract}
We study mirror symmetry of supermanifolds constructed as fermionic
extensions of compact toric varieties. We mainly discuss the case where the
linear sigma A-model contains as many fermionic fields as there are
$U(1)$ factors in the gauge group. In the
mirror super-Landau-Ginzburg  B-model, focus is on the bosonic structure
obtained after integrating out all the fermions. Our key observation is
that there is a relation between the super-Calabi-Yau
conditions of the A-model and quasi-homogeneity of the B-model,
and that the degree of the associated superpotential in the B-model
is given in terms of the determinant of the fermion charge matrix
of the A-model.
\end{abstract}

{\tt  KEYWORDS}: Mirror symmetry, supermanifolds, toric geometry.

\end{titlepage}

\newpage
\tableofcontents
\newpage

\section{Introduction}

Mirror symmetry underlies one of the most important and interesting
examples of string dualities, and provides a symmetry
between Calabi-Yau (CY) manifolds interpreted in terms of closed
topological string theories. More generally, the so-called A- and B-models
are related by mirror symmetry, as discussed below.
It has been realized, though, that rigid CY manifolds can
have mirror manifolds which are not themselves CY geometries.
An intriguing remedy is the introduction of CY {\em supermanifolds}
in these considerations \cite{Se,Sc}. It has thus been suggested that
mirror symmetry is between supermanifolds and manifolds alike,
and not just between bosonic manifolds.

It has been found recently that there is a correspondence
between the moduli space of holomorphic Chern-Simons theory on the
CY supermanifold $\mathbf{CP}^{3|4}$ and, self-dual, 
four-dimensional $N=4$ Yang-Mills theory \cite{W1,PS}. This may also
be related to the B-model of open  topological string theory
having $\mathbf{CP}^{3|4}$ as target space. Partly based on this
work, CY supermanifolds have subsequently attracted a great deal
of attention \cite{NV,AV1,KP,PW,Ahn,RW,Z}. It has been found, for
instance, that an A-model defined on the CY supermanifold
$\mathbf{CP}^{3|4}$ is a mirror of a B-model on a quadric
hypersurface in $\mathbf{CP}^{3|3}\times\mathbf{CP}^{3|3}$,
provided the K\"ahler parameter of $\mathbf{CP}^{3|4}$ approaches
infinity \cite{NV,AV1}. Following this observation, a possible
generalization of the A-model has been considered in which
fermionic coordinates with {\em different} weights are introduced
without changing the bosonic manifold $\mathbf{CP}^{3}$
\cite{PW,Ahn}.

The aim of the present work is to study mirror symmetry based
on a broad class of supermanifolds whose bosonic parts correspond to 
compact toric varieties. 
Important examples of such bosonic manifolds
are (weighted) projective spaces and products thereof.

Our analysis is based on the following scenario. The bosonic part
of the A-model is constructed as a $U(1)^{\otimes n}$ linear sigma
model whose target space is a toric variety. Adding a set of $f$
fermionic fields with charges given by an $n\times f$ matrix to
the sigma model, corresponds to extending the toric variety to a
supermanifold with $f$ Grassmannian coordinates. By extending the
T-duality prescription in \cite{AV1} on fermionic fields to cover
the {\em product} gauge group $U(1)^{\otimes n}$, we can obtain
the path integral description of the mirror super-Landau-Ginzburg
(super-LG) B-model. It initially involves $n$ delta functions
which may be integrated out to extract information on the
associated (super-)geometry. We shall focus on the bosonic
structure obtained after integrating out all the fermionic fields
in the B-model. Different patches may result depending on which
bosonic fields are integrated in the elimination of the delta
functions. We pay particular attention to the 'quadratic' case
where $f=n$, and consider generic values of the K\"ahler
parameters, that is, we do not restrict ourselves to simplifying
limits.

Our key observation in this set-up is that
the super-CY conditions of the A-model geometry are related to
quasi-homogeneity of the bosonic toric data of the B-model.
Furthermore, the degree of the
associated quasi-homogeneous superpotential in the B-model
is given in terms of the determinant
of the matrix of fermion charges in the A-model.
Details on this correspondence and the dependence on the
determinant will be provided in the main text.

After a brief summary of T-duality for fermionic coordinates or
fields in section 2, we discuss mirror symmetry of
supermanifolds in section 3. Our main result involving
the super-CY conditions and the determinant of fermion charges
is derived for products of weighted super-projective spaces.
We emphasize the situation for complex three-dimensional
projective spaces due to their relevance in string theory, cf. the
case of $\mathbf{CP}^{3|4}$ alluded to above. We also
relate our results based on $\mathbf{CP}^1\times\mathbf{CP}^{p-1}$
to superpotentials discussed in \cite{Sch}.
Section 4 concerns the extension to general toric varieties,
and we find that our key observation still holds.
The family $\{\tilde{F}_m\}$ of three-dimensional toric
varieties generalizing the projective spaces are used
as an illustration. A conclusion is presented in section 5.

\section{T-duality of fermionic fields}

In this section we review T-duality for fermionic coordinates 
(fields), and we do this by first recalling the bosonic case \cite{HV}.

To this end, we consider a linear sigma model described in terms
of the chiral fields ${\Phi}_i$, $i=1, \ldots,p$, with charges
$q_i$ under a $U(1)$ gauge symmetry \cite{W2}. The requirement of
conformal invariance of this system is equivalent to the CY
condition of the target space,
\be
 \sum_{i}q_{i}=0.
\ee
The vanishing condition for the potential energy density for the scalar
fields reads
\be
\label{dterm}
 \sum_{i}{q_{i}{|{\Phi}_i|}^{2}}=r
\ee 
where $r$ is a Fayet-Iliopoulos (FI) coupling constant which,
combined with the $U(1)$ $\theta$-angle, defines the complexified
K\"ahler parameter $t=r+i\theta$. Note in passing that eq.
(\ref{dterm}) corresponds to a local CY manifold.

Following \cite{HV,AV2, B}, the mirror model is obtained by
introducing a set of fields $\{Y_i\}$ dual to the set
$\{\Phi_i\}$, such that
\be
 \Re(Y_i)=|\Phi_i|^2,
\ee 
where $\Re(Y_i)$ denotes the real part of $Y_i$. The mirror
version of (\ref{dterm}) is
\be
\label{mirror}
 \sum_{i}q_{i}Y_i=t
\ee
and the corresponding superpotential in the associated
LG model reads
\be
\label{WmirrorY}
 W=\sum_{i} e^{-Y_i}.
\ee
Using the following field redefinitions
\be
  \hat{y}_{i}=e^{-Y_{i}},
\label{yY}
\ee
the superpotential becomes
\be
\label{Wmirrory}
 W=\sum_{i}\hat{y}_i,
\ee
subject to
\be
 \prod_i \hat{y}_i^{q_i}=e^{-t}.
\ee 
With multiple toric actions, $U(1)^{\otimes n}$, this extends
readily to
\be
 \prod_i \hat{y}_i^{q^a_i}=e^{-t^a},\ \ \ \ \ \ \ a=1\ldots,n.
\label{ta} 
\ee 
Here $t^a$ is the complexified K\"ahler parameter
associated to the $a$th $U(1)$ factor, while $q^a$ is the
charge vector with respect to the same $U(1)$ factor.

It has been shown recently that a similar analysis can be carried
out for {\em fermionic} fields as well, though with a different
rule for 'dualizing' the fields \cite{AV1}. For a system with
fermionic fields $\{\Psi_\al\}$ with charges $Q_\al$, and bosonic
fields $\{\Phi_i\}$ as above, the extended D-term constraint of eq.
(\ref{dterm}) reads
\be
 \sum_{i}{q_{i}{|{\Phi}_i|}^{2}}+\sum_{\alpha}{Q_{\alpha}
  {|{\Psi}_\alpha|}^{2}}=\Re(t).
\ee 
The condition for the associated super-variety to be a CY
supermanifold is given by
\be
 \sum_iq_i=\sum_\al Q_\al.
\label{CYA} 
\ee 
Under T-duality, the bosonic superfield ${\Phi}_i$
of the linear sigma model is still replaced by a dual superfield
$Y_i$. The fermionic superfield ${\Psi}_\alpha$, on the other
hand, is dualized by the triplet $(X_\al,\eta_\al,\chi_\al)$
\cite{AV1}, where the bosonic field $X_\al$ satisfies
\be
 \Re(X_\al)=-|\Psi_\al|^2.
\ee
The accompanying pair of fields, $\eta_\al$ and $\chi_\al$, are fermionic
and required to preserve the superdimension and hence
the total central charge under the symmetry. The corresponding
mirror super-LG model is given by the path integral
\bea
 {\cal Z}&=&\int
 \left({\prod_i dY_i}\right)
 \left({\prod_\alpha dX_\alpha d\eta_\alpha d\chi_\alpha}\right)
  \delta\left(\sum_{i}q_{i}Y_i-\sum_{\alpha}Q_{\alpha}X_\alpha-t\right)\nn
 &\times&
  \exp\left(\sum_{i}e^{-Y_i}+\sum_{\alpha}e^{-X_\alpha}
  (1+\eta_\alpha\chi_\alpha)\right).
\label{ZLG}
\eea

The objective in the following is to extend this analysis to a
linear sigma A-model with {\em product} gauge group $U(1)^{\otimes
n}$ and $f$ fermionic fields, and study the resulting mirror
B-model as defined by a generalization of (\ref{ZLG}). One may in
this case supplement the field redefinitions in (\ref{yY}) with
\be
 \hat{x}_\al=e^{-X_\al},\ \ \ \ \ \ \ \al=1,\ldots,f.
\label{xX}
\ee
The associated conditions (\ref{ta}) on the bosonic part of the
B-model superpotential then read
\be
  \prod_i \hat{y}_i^{q^a_i}=e^{-t^a}\prod_\al\hat{x}^{Q_\al^a},
  \ \ \ \ \ \ \ a=1,\ldots,n.
\label{yxta}
\ee
Focus will be on the toric data of the bosonic structure obtained
after eliminating the $n$ delta functions and integrating out the
$2f$ fermionic fields in the B-model path integral. We shall find that this
bosonic structure is described in terms of
the set of fermion charges in the A-model. In particular, the
super-CY conditions of the A-model extending (\ref{CYA}) turn out to be
related to quasi-homogeneity of the bosonic structure
of the B-model.

\section{Mirrors of super-projective spaces}

We recall that a general complex $p$-dimensional toric variety can
be expressed in the following form,
\be
 \mathbf{V}^p = \frac{\mathbb{C}^{p+n}\setminus S}{{\mathbb{C}^*}^n},
\label{Vpgen}
\ee
where the $n$ $\mathbb{C}^*$ actions are given by
\be
 {\mathbb{C}^*}^n: z_i \to \la^{q_i^a} z_i,\quad\ \   i=1,\ldots, p+n;
 \quad\ \  a=1,\ldots,n.
\ee 
In these expressions, the exponents ${q_i^a}$ are referred to as
charges and are assumed to be
integers. For each fixed $a$ they define a Mori vector in toric geometry. 
These vectors thus generalize the weight vector $w$ of the weighted
projective space ${\bf WP}^p_{(w_1,\ldots,w_{p+1})}$. The
subtracted part $S$ is a subset of $\mathbb{C}^{p+n}$ chosen by
triangulation. The variety $\mathbf{V}^p$ can be represented by a
toric diagram $\Delta(\mathbf{V}^p)$ spanned by $k=p+n$ vertices
$v_i$ in a $\mathbb{Z}^p$ lattice satisfying
\be
  \sum \limits _{i=1}^{p+n} q_i^a v_ i=0,\quad\ \ \  a=1,\ldots,n.
\ee 
It may be realized in terms of an $N=2$ linear sigma model,
where one considers a two-dimensional supersymmetric $N=2$ gauge
system with $U(1)^{\otimes n}$ gauge group and $p+n$ chiral fields
$\Phi_i$ with a charge matrix whose entries are $q_i^a$
\cite{W2}. In this way, and
up to $U(1)^{\otimes n}$ gauge transformations, the K\"ahler
manifold $\mathbf{V}^p$ corresponds to the minimum of the D-term
potential (D$^a=0$). That is,
\be
  \sum \limits _{i=1}^{p+n} q_i^a |\Phi_i|^2=r^a,
\ee
where the $r^a$'s are FI coupling parameters.

Here we consider the complex $p$-dimensional toric variety
defined by the following trivial fibration
\be
 \mathbf{WCP}^{p_1-1}_{(w_1^1,\ldots,w_{p_1}^1)}\times
  \mathbf{WCP}^{p_2-1}_{(w_1^2,\ldots,w_{p_2}^2)}\times\ldots\times
  \mathbf{WCP}^{p_n-1}_{(w_1^n,\ldots,w_{p_n}^n)}
\label{WCP} 
\ee
where $p=\sum_{a=1}^n(p_a-1)$. It admits a $U(1)^{\otimes n}$
sigma-model description in terms of the $p+n$ bosonic fields
\be
 \{\Phi_1^1,\ldots,\Phi_{p_1}^1;\Phi_1^2,\ldots,\Phi_{p_2}^2;
  \ldots;\Phi_1^n,\ldots,\Phi_{p_n}^n\}
\label{Phi}
\ee
with charge vectors
\bea
 q^1&=&(w_1^1,\ldots,w_{p_1}^1;0,\ldots,0;\ldots;0,\ldots,0),\nn
 q^2&=&(0,\ldots,0;w_1^2,\ldots,w_{p_2}^2;0,\ldots,0;\ldots;0,\ldots,0),
  \nn
    &\vdots&\nn
 q^n&=&(0,\ldots,0;\ldots;0,\ldots,0;w_1^n,\ldots,w_{p_n}^n).
\label{q}
\eea
A toric variety like (\ref{WCP}) is compact if all the
charges (\ref{q}) are positive (or negative) integers.
Its associated sigma model is a solution of the D-term constraints
\be
 \sum_{i=1}^{p_a}w_{i}^a|\Phi_i^a|^2=\Re(t^a),\ \ \ \ \ \ \
   a=1,\ldots,n
\label{Dmn} 
\ee 
where $t^a$ is the complexified K\"ahler parameter
associated to the $a$th factor in (\ref{WCP}). All
weights $w_i^a$ are assumed non-vanishing. By convention for
weighted projective spaces, the greatest common divisor of the
weights $w_i^a$ for a given $a$ is one.

The objective now is to consider a fermionic extension of the
manifold (\ref{WCP}), thus turning it into a (weighted)
super-projective space and discuss its mirror companion. Our
approach may be seen as an illustration and an extension of the
previous section by taking into account the product structure of
(\ref{WCP}) with its enlarged symmetry.

Adding $f$ Grassmann coordinates to (\ref{WCP}) corresponds to
supplementing the bosonic sigma model, described by (\ref{Phi}),
by $f$ fermionic fields,
\be
 \{\Psi_\al,\ \al=1,\ldots,f\},
\label{Psi} 
\ee 
with charges $Q_\al^a$. The full spectrum of
$U(1)^{\otimes n}$ charge vectors thus becomes 
\bea
 q'^1&=&( q^1\ |\
  Q_1^1,\ldots,Q_f^1),\nn
 q'^2&=&(q^2
  \ |\ Q_1^2,\ldots,Q_f^2),\nn
    &\vdots&\nn
 q'^n&=&(q^n\ |\
  Q_1^n,\ldots,Q_f^n),
\label{qQ}
\eea
while the extended D-term constraints of this A-model (cf. (\ref{Dmn})) read
\be
 \sum_{i=1}^{p_a}w_i^a |\Phi_i^a |^2+\sum_{\al=1}^fQ_\al^a |\Psi_\al |^2
 =\Re(t^a),\ \ \ \ \ \ \ a=1,\ldots,n.
\label{Ds}
\ee

There is an abundance of possible fermionic extensions following this
prescription. It
may be limited, though, by imposing the super-CY conditions (\ref{CYA}):
\be
 0=\sum_iq_i^a-\sum_\al Q_\al^a
  =\sum_{i=1}^{p_a}w_i^a-\sum_{\al=1}^fQ_\al^a,\ \ \ \ \ \ \
   a=1,\ldots,n.
\label{sCY} 
\ee 
We shall initially refrain from doing this since one of our key
observations will be that these conditions are related
to quasi-homogeneity of a bosonic structure of the B-model geometry.
This new correspondence between two mirror supermanifolds
will be addressed below.

Before proceeding, we recall that the charge vectors (\ref{q})
and D-term constraints (\ref{Dmn}) of the bosonic A-model
correspond to a sigma-model realization of the toric variety
(\ref{WCP}). This extends readily to general
toric varieties as defined by (\ref{Vpgen}), and we shall have more
to say about this in section 4. Here we wish to point out
that in a similar fashion the expressions (\ref{qQ}) and (\ref{Ds}) 
may be seen as corresponding to an $N=2$ sigma-model
realization of a {\em fermionic} extension of a toric variety.
A {\em super-variety} 
\be
 \mathbf{V}^{p|f}=\frac{\mathbb{C}^{p+n|f}\setminus S}{{\mathbb{C}^*}^n} 
\label{Vpf}
\ee
is thereby represented by a toric super-diagram spanned by 
$(p+n)$ vertices $v_i$ and $f$ vertices $v_\alpha$ 
in a superlattice $\mathbb{Z}^{p|f}$, and constrained as 
\be
 \sum_{i=1}^{p+n} q^{a}_{i}v_i -\sum_{\alpha=1}^f
  Q^{a}_{\alpha}v_{\alpha}=0,\ \ \ \ \ \ \ \ a=1,\ldots,n.
\label{pf}
\ee
Here we have used the notation $\mathbb{C}^{p+n|f}$ to
indicate a fermionic extension of $\mathbb{C}^{p+n}$.
We suggest to refer to this fermionic extension of toric geometry
as toric super-geometry. It is seen that CY supermanifolds
are defined naturally in toric super-geometry.

According to the T-duality outlined in the previous section, the
mirror B-model is now obtained by replacing the field $\Phi_i^a$ by a
superfield $Y_i^a$, while the fermionic field $\Psi_\al$ is
dualized by the triplet $(X_\al,\eta_\al,\chi_\al)$. Applying the
mirror symmetry transformation to the A-model above thus results
in a B-model in terms of a super-LG mirror model given by the path
integral \bea
 {\cal Z}&=&\int\left(\prod_{i=1}^{p_1}dY_i^1\right)
  \ldots\left(\prod_{i=1}^{p_n}dY_i^n\right)
  \left(\prod_{\al=1}^fdX_\al d\eta_\al d\chi_\al\right)\nn
 &\times&\delta\left(\sum_{i=1}^{p_1}w_i^1Y_i^1
  -\sum_{\al=1}^fQ_\al^1X_\al-t^1\right)
  \times\ldots\times\delta\left(\sum_{i=1}^{p_n}w_i^nY_i^n
  -\sum_{\al=1}^fQ_\al^nX_\al-t^n\right)\nn
 &\times&\exp\left(\sum_{a=1}^n\sum_{i=1}^{p_a}
  e^{-Y_i^a}+\sum_{\al=1}^fe^{-X_\al}(1+\eta_\al\chi_\al)\right).
\label{Zmn}
\eea

To extract information on the B-model (super-)geometry, one would naturally
wish to integrate out the $n$ delta functions. In this paper, we shall
focus on
the 'quadratic' case where $n=f$ and subsequently choose to integrate out the
bosonic fields $X_\al$, $\al=1,\ldots,f=n$.
We intend to address elsewhere \cite{Work}
the situations where $f\neq n$ or where the elimination of the
delta functions may involve integrating out some of the fields $Y_i$.
As already mentioned, we are here interested in the bosonic structure
arising after integrating out all the $2f$ fermionic fields.

To illustrate the construction, focus here will be on the situation
where $n=2$, $p_1=2$ and $p_2=3$, while  in section 4 we
shall report on the case based on a general toric variety with $f=n$.
For now, our chosen A-model scenario thus corresponds to a fermionic
extension of the complex three-dimensional variety
$\mathbf{WCP}^{1}_{(w_1^1,w_2^1)}\times
\mathbf{WCP}^{2}_{(w_1^2,w_2^2,w_3^2)}$ which we shall
assume is compact.
A particular example is provided by $\mathbf{CP}^{1}\times\mathbf{CP}^{2}$
and corresponds to a trivial fibration of $\mathbf{CP}^{1}$
over the base space $\mathbf{CP}^{2}$.
This manifold is sometimes denoted
$\tilde F_0$ and has been used in the construction of
real four-dimensional $N=1$ models obtained from
F-theory compactification on elliptic CY fourfolds \cite{BM}.
We shall have more to say about the infinite family of complex
three-dimensional manifolds
$\tilde F_m$, as it appears as a particular subclass of the general study
in section 4.

\subsection{Mirrors of fermionic extensions of
  $\mathbf{WCP}^{1}_{(w_1^1,w_2^1)}\times
   \mathbf{WCP}^{2}_{(w_1^2,w_2^2,w_3^2)}$}

We are here considering the case with $f=n=2$.
The integral over the four fermionic fields $\eta_1$, $\eta_2$,
$\chi_1$ and $\chi_{2}$ appearing in the super-LG B-model (\ref{Zmn})
produces a simple expression in $X_{1}$ and $X_2$:
\be
 \int\left(\prod_{\al=1}^2d\eta_\al d\chi_\al\right)
  \exp\left(e^{-X_1}(1+\eta_1\chi_1)+e^{-X_2}(1+\eta_2\chi_2)
  \right)=e^{-X_1}e^{-X_2}\exp\left(e^{-X_1}+e^{-X_2}\right).
\label{intetachi}
\ee
Solving the delta-function constraints amounts to solving the two
linear equations
\bea
 Q_1^1X_1+Q_2^1X_2&=&w_1^1Y_1^1+w_2^1Y_2^1-t^1,\nn
 Q_1^2X_1+Q_2^2X_2&=&w_1^2Y_1^2+w_2^2Y_2^2+w_3^2Y_3^2-t^2.
\label{two}
\eea
There is a unique solution for $X_1$ and $X_2$ provided the determinant
\be
 D=Q_1^1Q_2^2-Q_2^1Q_1^2
\label{DQQ}
\ee
is non-vanishing. For vanishing determinant, the equations
(\ref{two}) would impose linear relations
among the fields $\{Y_i^a\}$. We shall assume that
$D\neq0$, in which case the path integral (\ref{Zmn}) reduces to
\bea
 {\cal Z}&\propto&\int\left(\prod_{i=1}^2dY_i^1\right)
  \left(\prod_{i=1}^3dY_i^2\right)
  e^{\frac{Q_1^2-Q_2^2}{D}\sum_{i=1}^2w_i^1Y_i^1
   -\frac{Q_1^1-Q_2^1}{D}\sum_{i=1}^3w_i^2Y_i^2}\nn
 &\times&\exp\left(\sum_{i=1}^2e^{-Y_i^1}+\sum_{i=1}^3e^{-Y_i^2}\right)
  \exp\left(e^{-\frac{Q_2^2}{D}\left[\sum_{i=1}^2w_i^1Y_i^1-t^1\right]
  +\frac{Q_2^1}{D}\left[\sum_{i=1}^3w_i^2Y_i^2-t^2\right]}\right)\nn
&\times&\exp\left(e^{\frac{Q_1^2}{D}\left[\sum_{i=1}^2w_i^1Y_i^1-t^1\right]
  -\frac{Q_1^1}{D}\left[\sum_{i=1}^3w_i^2Y_i^2-t^2\right]}\right).
\label{Ztwo}
\eea

Our current objective is to extract information on the geometry
associated to the path integral (\ref{Ztwo}).
This may be achieved naively if one can make field redefinitions
turning the path integral into the form
\be
 {\cal Z}\ \simeq \ \int\left(\prod_k^\ell d\varphi_k\right) 
  e^{-W(\{\varphi_k\})}.
\label{W}
\ee
The functional expression $W$ is then referred to as the superpotential,
and its vanishing condition, $W=0$, provides an algebraic
equation in terms of $\{\varphi_k\}$. For it to correspond to a conventional
LG theory, it should be quasi-homogeneous in the sense that
\be
 W(\{\la^{w_k}\varphi_k\})=\la^dW(\{\varphi_k\}),
\label{Wla}
\ee
where the integers $w_k$, $k=1,\ldots,\ell$, indicate the scaling property
of the fields $\{\varphi_k\}$, while $d$ denotes the degree of the
superpotential. The vanishing condition $W=0$ thus corresponds to a
hypersurface in the weighted projective space
$\mathbf{WP}^{\ell-1}_{(w_1,\ldots,w_\ell)}$.

Motivated by this and with reference to (\ref{Ztwo}), we thus introduce
the new fields $\{y_i^a\}$, related to $\{Y_i^a\}$ by
\bea
 y_i^1&=&e^{\frac{Q_1^2-Q_2^2}{D}w_i^1Y_i^1}  \ \ \ \ \ \ \ i=1,2,\nn
 y_i^2&=&e^{\frac{Q_2^1-Q_1^1}{D}w_i^2Y_i^2}  \ \ \ \ \ \ \ i=1,2,3.
\label{y} 
\eea 
For these mappings to be sensible and do the intended job, 
we must assume that $Q_1^1\neq Q_2^1$ and $Q_1^2\neq Q_2^2$.
These assumptions may therefore be interpreted as an
initial requirement for (\ref{Ztwo}) to correspond to
a super-LG model. It is noted that they are neither necessary nor
sufficient conditions for the non-vanishing of the determinant $D$.
The path integral now reads
\bea
 {\cal Z}&\propto&\int\left(\prod_{i=1}^2dy_i^1\right)
  \left(\prod_{i=1}^3dy_i^2\right)\exp\left(\sum_{i=1}^2
  (y_i^1)^{\frac{D}{w_i^1(Q_2^2-Q_1^2)}}+
  \sum_{i=1}^3(y_i^2)^{\frac{D}{w_i^2(Q_1^1-Q_2^1)}}\right.\nn
 &&\left.+e^{\frac{Q_2^2t^1-Q_2^1t^2}{D}}
  (y_1^1y_2^1)^{\frac{Q_2^2}{Q_2^2-Q_1^2}}
  (y_1^2y_2^2y_3^2)^{\frac{Q_2^1}{Q_2^1-Q_1^1}}
  +e^{\frac{Q_1^1t^2-Q_1^2t^1}{D}}(y_1^1y_2^1)^{\frac{Q_1^2}{Q_1^2-Q_2^2}}
   (y_1^2y_2^2y_3^2)^{\frac{Q_1^1}{Q_1^1-Q_2^1}}\right),
\label{Yint}
\eea
and the vanishing of the superpotential turns into the algebraic equation
\bea
 0&=&\sum_{i=1}^2
  (y_i^1)^{\frac{D}{w_i^1(Q_2^2-Q_1^2)}}+
  \sum_{i=1}^3(y_i^2)^{\frac{D}{w_i^2(Q_1^1-Q_2^1)}}\nn
 &+&e^{\frac{Q_2^2t^1-Q_2^1t^2}{D}}
  (y_1^1y_2^1)^{\frac{Q_2^2}{Q_2^2-Q_1^2}}
  (y_1^2y_2^2y_3^2)^{\frac{Q_2^1}{Q_2^1-Q_1^1}}
  +e^{\frac{Q_1^1t^2-Q_1^2t^1}{D}}(y_1^1y_2^1)^{\frac{Q_1^2}{Q_1^2-Q_2^2}}
   (y_1^2y_2^2y_3^2)^{\frac{Q_1^1}{Q_1^1-Q_2^1}}.
\label{homD}
\eea
In order to determine the appropriately associated weighted projective space,
we consider the exponents of $y_i^1$ and $y_i^2$ in the two
sums. Let
\be
 g=gcd(w_1^1(Q_2^2-Q_1^2),w_2^1(Q_2^2-Q_1^2),
  w_1^2(Q_1^1-Q_2^1),w_2^2(Q_1^1-Q_2^1),w_3^2(Q_1^1-Q_2^1))
\label{g}
\ee
denote the greatest common divisor of the denominators of these
five exponents. The weighted projective space is then given by
\be
 \mathbf{WP}^{4}_{(\frac{w_1^1(Q_2^2-Q_1^2)}{g},
  \frac{w_2^1(Q_2^2-Q_1^2)}{g},
  \frac{w_1^2(Q_1^1-Q_2^1)}{g},\frac{w_2^2(Q_1^1-Q_2^1)}{g},
   \frac{w_3^2(Q_1^1-Q_2^1)}{g})}
 (y_1^1,y_2^1,y_1^2,y_2^2,y_3^2).
\label{WP}
\ee
The superpotential (\ref{homD}) is now quasi-homogeneous
provided the remaining, K\"ahler-dependent terms also have
degree $D/g$, where $D$ is the determinant (\ref{DQQ}).
This is the case {\em exactly provided}
the super-CY conditions (\ref{sCY}) are satisfied, which
in the present example, reduce to
\bea
 w_1^1+w_2^1&=&Q_1^1+Q_2^1,\nn
 w_1^2+w_2^2+w_3^2&=&Q_1^2+Q_2^2.
\label{sCY6}
\eea
This illustrates the announced correspondence between the super-CY
conditions of the A-model and quasi-homogeneity of the B-model,
and that the degree of the associated superpotential is given in terms of
the determinant of the fermion charge matrix of the A-model.
This is a new relation between a pair of mirror supermanifolds.

It is noted that the LG superpotential given by (\ref{homD}) in
general is not polynomial, and may include non-integer powers of
the coordinates  $y$. Let us examine when it {\em does} correspond
to a polynomial for {\em generic} $t^1$ and $t^2$. That is, the
two terms multiplied by the exponential expressions in the
K\"ahler parameters are present. Particular correlated limits of
these parameters could eliminate these terms and the conditions
for polynomial behaviour accompanying them. The strong correlation
with the super-CY condition, which has been derived for generic
K\"ahler parameters, may therefore be lost in certain limits.

For (\ref{homD}) to be polynomial, we must require that all powers
are non-negative integers. It follows from a comparison of the
powers of $(y_1^1y_2^1)$ in the two K\"ahler-dependent terms that
either $Q_2^2$ or $Q_1^2$ must vanish. Likewise, from the powers
of $(y_1^2y_2^2y_3^2)$ we find that either $Q_1^1$ or $Q_2^1$
vanishes. Since $D\neq0$, we then have the two possibilities:
(I) $Q_1^2=Q_2^1=0$, or (II) $Q_1^1=Q_2^2=0$. From the
K\"ahler-independent terms it then follows that (I)
$Q_1^1/w_i^1\in\mathbb{Z}_>$ and $Q_2^2/w_i^2\in\mathbb{Z}_>$, or
(II) $Q_2^1/w_i^1\in\mathbb{Z}_>$ and
$Q_1^2/w_i^2\in\mathbb{Z}_>$. {} From imposing the super-CY
condition (or quasi-homogeneity) as well, it follows in both
cases, (I) and (II), that $w_1^1=w_2^1$ and
$(w_1^2,w_2^2,w_3^2)\in \{P(k,k,k),P(k,k,2k),P(k,2k,3k)\}$, where
$P$ denotes a permutation and $k$ is a non-vanishing integer. As
discussed above, the greatest common divisor is conventionally
one, limiting the possible values to $k=\pm1$. Note that in these
considerations, the sign of the determinant is related to the
signs of the weight vectors $(w_1^1,w_2^1)$ and
$(w_1^2,w_2^2,w_3^2)$ in the A-model. A homogeneous and polynomial
structure thus arises in the B-model when the A-model is based on
$\mathbf{CP}^1\times\mathbf{WP}^2_{(-1,-3,-2)}$, for example,
provided the fermionic extension is governed by $Q_1^1=Q_2^2=0$,
$Q_2^1=2$ and $Q_1^2=-6$. The degree of the polynomial is then
$D/2=6$, and it describes a hypersurface in the compact, weighted
projective space
$\mathbf{WP}^4_{(3,3,1,3,2)}(y_1^1,y_2^1,y_1^2,y_2^2,y_3^2)$. A
simpler example arises when choosing to base the A-model on
$\mathbf{CP}^1\times\mathbf{CP}^2$ with fermionic extension given
by $Q_1^1=2$, $Q_2^2=3$ and $Q_1^2=Q_2^1=0$. The bosonic structure
of the B-model is then described by a polynomial hypersurface of
degree 6 in
$\mathbf{WP}^4_{(3,3,2,2,2)}(y_1^1,y_2^1,y_1^2,y_2^2,y_3^2)$.

There are several possible generalizations of the above analysis.
We shall discuss some of them below,
while others will be addressed elsewhere \cite{Work}.

\subsection{Mirrors of fermionic extensions of $\mathbf{CP}^1\times
 \mathbf{CP}^{p-1}$}

A first and simple generalization is to consider a sigma model
whose target space is a fermionic extension of
$\mathbf{CP}^1\times\mathbf{CP}^{p-1}$, $p\geq2$. This corresponds
to a $U(1)\otimes U(1)$ gauge theory with $p+2$ bosonic fields
$\Phi_i$ and (in the quadratic case where $n=f$) two fermionic
fields $\Psi_\al$ with charges
\be
\label{ccharge}
 q'^{1}=(1,1,0,0,\dots,0\ |\ Q^1_1,Q^1_2 ), \qquad \ \
  q'^{2}=(0,0,1,1,\dots,1\ |\ Q^2_1,Q^2_2).
\ee
The D-term constraints of this A-model are given by
\bea
 |\Phi_1|^2+|\Phi_2|^2+Q_1^1|\Psi_1|^2+Q_2^1|\Psi_2|^2&=&\Re(t^1),\nn
 \sum_{i=3}^{p+2}|\Phi_i|^2+Q_1^2|\Psi_1|^2+Q_2^2|\Psi_2|^2
  &=&\Re(t^2),
\label{D1n}
\eea
while the super-CY conditions read
\be
 Q_1^1+Q_2^1=2,\ \ \ \ \ \ \ Q_1^2+Q_2^2=p.
\label{pCY}
\ee

Now, following the prescription above, we introduce
the field redefinitions
\bea
 y_i&=&e^{\frac{Q_1^2-Q_2^2}{D}Y_i},\ \ \ \ \ \ \ i=1,2,\nn
 y_i&=&e^{\frac{Q_2^1-Q_1^1}{D}Y_i},\ \ \ \ \ \ \ i=3,\ldots,p+2,
\label{yp}
\eea
resulting in a B-model superpotential whose vanishing condition
is given by
\bea
 0&=&y_1^{\frac{D}{Q_2^2-Q_1^2}}+y_2^{\frac{D}{Q_2^2-Q_1^2}}
  +\sum_{i=3}^{p+2}y_i^{\frac{D}{Q_1^1-Q_2^1}}\nn
  &+&e^{\frac{Q_2^2t^1-Q_2^1t^2}{D}}(y_1y_2)^{\frac{Q_2^2}{Q_2^2-Q_1^2}}
   (y_3\ldots y_{p+2})^{\frac{Q_2^1}{Q_2^1-Q_1^1}}
   +e^{\frac{Q_1^1t^2-Q_1^2t^1}{D}}(y_1y_2)^{\frac{Q_1^2}{Q_1^2-Q_2^2}}
   (y_3\ldots y_{p+2})^{\frac{Q_1^1}{Q_1^1-Q_2^1}}.
\label{ppot}
\eea
With
\be
 D=Q_1^1Q_2^2-Q_1^2Q_2^1,\ \ \ \ \ \ g=gcd(Q_2^2-Q_1^2,Q_1^1-Q_2^1)
\label{pg}
\ee
and the super-CY conditions (\ref{pCY}) imposed,
we find that (\ref{ppot}) corresponds to a hypersurface of degree
$D/g$ in the weighted projective space
$\mathbf{WP}^{p+1}_{(\frac{Q_2^2-Q_1^2}{g},\frac{Q_2^2-Q_1^2}{g},
 \frac{Q_1^1-Q_2^1}{g},\ldots,\frac{Q_1^1-Q_2^1}{g})}(y_1,\ldots,y_{p+2})$.
The conditions for the superpotential to be polynomial are
(I) $Q_1^2=Q_2^1=0$, or (II) $Q_1^1=Q_2^2=0$.
It is noted that with the super-CY conditions imposed, $g=1$
for $p$ odd, while $g=2$ for $p$ even.

Let us analyze the two options for a polynomial superpotential,
namely (I) or (II). It turns out that in either case,
the polynomial and quasi-homogeneous superpotential reads
\be
 0=y_1^2+y_2^2+\sum_{i=3}^{p+2}y_i^p+e^{t^1/2}y_1y_2+
  e^{t^2/p}y_3\ldots y_{p+2}
\label{pI}
\ee
and describes a hypersurface of degree $2p/g$ in
\bea
 \mathbf{WP}^{p+1}_{(p,p,2,\ldots,2)}(y_1,\ldots,y_{p+2}),&&\ \ \ \ \ \ \
  p\ {\rm odd},\nn
 \mathbf{WP}^{p+1}_{(\frac{p}{2},\frac{p}{2},1,\ldots,1)}
  (y_1,\ldots,y_{p+2}),&&\ \ \ \ \ \ \ p\ {\rm even}.
\label{wp}
\eea
That is, the degree is $2p$ for $p$ odd, and $p$ for $p$ even.
This infinite family of weighted projective spaces has already
appeared in the literature \cite{Sch}. There\footnote{Here the
family is labeled by $p$, $p\geq2$, which in \cite{Sch} is
denoted $n+1$.} it is discussed that a quasi-homogeneous
hypersurface of degree $2p$ (for $p$ odd) or $p$ (for $p$ even)
in the space (\ref{wp}) should be of relevance to mirror symmetry
of higher-dimensional manifolds. This is confirmed here since we
have found that such hypersurfaces correspond to bosonic
structures of supermanifolds which are mirror partners to
fermionic extensions of $\mathbf{CP}^1\times\mathbf{CP}^{p-1}$.

\section{Mirrors of fermionic extentions of toric varieties}

Now we extend our study of super-projective spaces to fermionic
extensions of general toric varieties. With reference to the
description at the beginning of section 3, in particular, the
A-model is based on the $p+n$ bosonic fields $\Phi_i$ and the $f$
fermionic fields $\Psi_\al$ with $U(1)^{\otimes n}$ charges
\be
 q'^a=(q^a\ |\ Q^a)=(q_1^a,\ldots,q_{p+n}^a\ |\ Q_1^a,\ldots, Q_f^a),
  \ \ \ \ \ \ \ a=1,\ldots,n.
\label{qgen} 
\ee 
The extended D-term constraint equations of the present
A-model reads
\be
 \sum_{i=1}^{p+n}q_i^a|\Phi_i|^2+\sum_{\al=1}^fQ_\al^a|\Psi_\al|^2
 =\Re(t^a),\ \ \ \ \ \ \ a=1,\ldots,n.
\label{Dgen}
\ee

The associated mirror B-model is obtained in the same way as
above, and the super-LG path integral becomes 
\bea
 {\cal Z}&=&\int\left(\prod_{i=1}^{p+n}dY_i\right)
  \left(\prod_{\al=1}^fdX_\al d\eta_\al d\chi_\al\right)\nn
 &\times&\delta\left(\sum_{i=1}^{p+n}q_i^1Y_i
  -\sum_{\al=1}^fQ_\al^1X_\al-t^1\right)
  \times\ldots\times\delta\left(\sum_{i=1}^{p+n}q_i^nY_i
  -\sum_{\al=1}^fQ_\al^nX_\al-t^n\right)\nn
 &\times&\exp\left(\sum_{i=1}^{p+n}
  e^{-Y_i}+\sum_{\al=1}^fe^{-X_\al}(1+\eta_\al\chi_\al)\right).
\label{Zgen} 
\eea 
Following the same procedure as before, we
integrate out the $2f$ fermionic fields yielding
\be
 \int\left(\prod_{\al=1}^fd\eta_\al d\chi_\al\right)
   \exp\left(\sum_{\al=1}^fe^{-X_\al}(1+\eta_\al\chi_\al)\right)
  =\left(\prod_{\al=1}^fe^{-X_\al}\right)
  \exp\left(\sum_{\al=1}^fe^{-X_\al}\right).
\label{fgen} 
\ee 
Focusing on the interesting situation where $n=f$ and where
the delta functions appearing in (\ref{Zgen}) are eliminated by
integrating out the fields $X_\al$, the set of linear equations
expressing the $X$ fields in terms of the $Y$ fields is given by
\be
 \sum_{\al=1}^fQ_\al^aX_\al=\sum_{i=1}^{p+f}q_i^aY_i-t^a,
   \ \ \ \ \ \ a=1,\ldots,f.
\label{Xgen} 
\ee 
There is a unique solution for this system of
equations provided the determinant of the quadratic $f \times f$
matrix of fermion charges $Q$,
\be
 D=\det(Q),
\label{Detgen} 
\ee 
is non-vanishing. We shall assume this. The
solution to (\ref{Xgen}) is then given in terms of the inverse
matrix $Q^{-1}$ as it may be written
\be
 X_\al=\sum_{a=1}^f(Q^{-1})_\al^a
   \left(\sum_{i=1}^{p+f}q_i^aY_i-t^a\right).
\label{Xsol}
\ee
Note that if $a$
is interpreted as the row index in $Q_\al^a$, as in (\ref{qQ}), it corresponds
to the column index in $(Q^{-1})^a_\al$.
After integrating out these fields, the path integral (\ref{Zgen})
is free of delta functions:
\bea
 {\cal Z}&\propto&\int\left(\prod_{i=1}^{p+f}dY_i\right)
  \exp\left(-\sum_{\al=1}^f\sum_{a=1}^f\sum_{i=1}^{p+f}
   (Q^{-1})_\al^aq_i^aY_i\right)\nn
 &\times&\exp\left(\sum_{i=1}^{p+f}e^{-Y_i}+\sum_{\al=1}^f
  \exp\left(-\sum_{a=1}^f(Q^{-1})_\al^a
   \left(\sum_{i=1}^{p+f}q_i^aY_i-t^a\right)\right)\right).
\label{Zgen2}
\eea

In order to extract information on the underlying geometry,
we again follow the prescription outlined in section 3.
We therefore introduce the field redefinitions
\be
 y_i=\exp\left(-\sum_{\al=1}^f\sum_{a=1}^f
   (Q^{-1})_\al^aq_i^aY_i\right),
\label{ygen}
\ee
and require that
\be
 \sum_{\al=1}^f\sum_{a=1}^f
   (Q^{-1})_\al^aq_i^a\neq0,\ \ \ \ \ \ \ i=1,\ldots,p+f.
\label{neq}
\ee
This ensures, in particular, that the superpotential
may be written as a finite sum of products of powers of the
fields. The path integral (\ref{Zgen2}) now reads
\bea
 {\cal Z}&\propto&\int\left(\prod_{i=1}^{p+f}dy_i\right)
  \exp\left(\sum_{i=1}^{p+f}
   y_i^{1/\{\sum_{\al=1}^f\sum_{a=1}^f(Q^{-1})_\al^aq_i^a\}}
  \right.\nn
 &&\left. +\sum_{\al=1}^fe^{\sum_{c=1}^f(Q^{-1})_\al^ct^c}
  \prod_{i=1}^{p+f}y_i^{\{\sum_{a=1}^f(Q^{-1})_\al^aq_i^a\}/\{
   \sum_{\beta=1}^f\sum_{b=1}^f(Q^{-1})_\beta^bq_i^b\}}\right).
\label{Zygen}
\eea
The vanishing of the superpotential thus defined may be
characterized by a hypersurface in the weighted projective space
\be
 \mathbf{WP}^{p+n-1}_{(
  \frac{D\sum_{\al=1}^f\sum_{a=1}^f(Q^{-1})_\al^aq_1^a}{g},
   \ldots,\frac{D\sum_{\al=1}^f\sum_{a=1}^f(Q^{-1})_\al^aq_{p+n}^a}{g})}
    (y_1,\ldots,y_{p+n}),
\label{WPpn}
\ee
where we have introduced the parameter
\be
 g=gcd(D\sum_{\al=1}^f\sum_{a=1}^f(Q^{-1})_\al^aq_1^a,
   \ldots,D\sum_{\al=1}^f\sum_{a=1}^f(Q^{-1})_\al^aq_{p+n}^a).
\label{ggen}
\ee
The hypersurface is given by the algebraic equation
\be
 0=\sum_{i=1}^{p+f}
   y_i^{1/\{\sum_{\al=1}^f\sum_{a=1}^f(Q^{-1})_\al^aq_i^a\}}
   +\sum_{\al=1}^fe^{\sum_{c=1}^f(Q^{-1})_\al^ct^c}
  \prod_{i=1}^{p+f}y_i^{\{\sum_{a=1}^f(Q^{-1})_\al^aq_i^a\}/\{
   \sum_{\beta=1}^f\sum_{b=1}^f(Q^{-1})_\beta^bq_i^b\}}.
\label{0}
\ee
Note that the factors of the determinant $D$ (\ref{Detgen})
in the definition of
the weights in (\ref{WPpn}) are required in general to ensure that the
weights are integers.
The expression (\ref{0}) is seen to be quasi-homogeneous provided
\be
 \sum_{i=1}^{p+f}q_i^a=\sum_{\al=1}^fQ_\al^a,\ \ \ \ \ \ a=1,\ldots,f
\label{CYgen}
\ee
which are the super-CY conditions of the original fermionic
extension of the projective variety in the A-model. The degree
of the superpotential is then given by $D/g$.
This provides the most general version presented here
of the new correspondence
between two supermanifolds paired by mirror symmetry.

The question of when the superpotential is
polynomial is more complicated in this general case than
in the projective example in section 3. It is beyond the scope
of the present work to attempt such a classification, though
we intend to address it elsewhere \cite{Work}.

Instead, let us point out that the family of complex three-dimensional
toric varieties $\tilde{F}_m$, $m\geq0$, is covered by our analysis.
That is, one may start with an A-model constructed as a fermionic
extension of the toric variety $\tilde{F}_m$. It is characterized
by the charge vectors
\bea
 q'^1&=&(1,1,0,0,m\ |\ Q_1^1,Q_2^1)\nn
 q'^2&=&(0,0,1,1,1\ |\ Q_1^2,Q_2^2)
\label{F} 
\eea 
with respect to the gauge group $U(1)^{\otimes 2}$.
Imposing the super-CY conditions yields
\be
 Q_1^1+Q_2^1=m+2,\ \ \ \ \ \ \ Q_1^2+Q_2^2=3.
\label{FCY}
\ee
The superpotential (\ref{0}) reduces to
\bea
 0&=& y_1^{\frac{D}{Q_2^2-Q_1^2}} +  y_2^{\frac{D}{Q_2^2-Q_1^2}} +
   y_3^{\frac{D}{Q_1^1-Q_2^1}} +  y_4^{\frac{D}{Q_1^1-Q_2^1}} +
  y_5^{\frac{D}{Q_1^1-Q_2^1+m(Q_2^2-Q_1^2)}}\nn
 &+&e^{\frac{Q_2^2t^1-Q_2^1t^2}{D}}(y_1y_2)^{\frac{Q_2^2}{Q_2^2-Q_1^2}}
  (y_3y_4)^{\frac{Q_2^1}{Q_2^1-Q_1^1}}
   y_5^{\frac{Q_2^1-mQ_2^2}{Q_2^1-Q_1^1-m(Q_2^2-Q_1^2)}}\nn
  &+&e^{\frac{Q_1^1t^2-Q_1^2t^1}{D}}(y_1y_2)^{\frac{Q_1^2}{Q_1^2-Q_2^2}}
  (y_3y_4)^{\frac{Q_1^1}{Q_1^1-Q_2^1}}
   y_5^{\frac{Q_1^1-mQ_1^2}{Q_1^1-Q_2^1+m(Q_2^2-Q_1^2)}}
\label{Fm0}
\eea
and corresponds to a hypersurface in (\ref{WPpn}) which
now reads
\be
 \mathbf{WP}^{4}_{(\frac{Q_2^2-Q_1^2}{g}, \frac{Q_2^2-Q_1^2}{g},
  \frac{Q_1^1-Q_2^1}{g},
  \frac{Q_1^1-Q_2^1}{g}, \frac{Q_1^1-Q_2^1+m(Q_2^2-Q_1^2)}{g})}
  (y_1,y_2,y_3,y_4,y_5),
\label{FmWP}
\ee
where
\be
 g=gcd(Q_2^2-Q_1^2,Q_1^1-Q_2^1,Q_1^1-Q_2^1+m(Q_2^2-Q_1^2)
   =gcd(Q_2^2-Q_1^2,Q_1^1-Q_2^1).
\ee 
The degree of the superpotential is $D/g$. A simple adaptation
of the discussion of the polynomial behaviour of the
superpotential (\ref{homD}), reveals that in order for (\ref{Fm0})
to be a homogeneous polynomial, we again have the two cases (I)
and (II). In case (I), for example, where $Q_1^2=Q_2^1=0$, it
follows from a comparison of the powers of $y_5$ in the two
K\"ahler-dependent terms that one of the three entities $Q_1^1$,
$Q_2^2$ or $m$ must vanish. Since $D\neq0$ we see that $m=0$. A
similar argument applies to case (II). We may thus conclude that
the only $\tilde{F}_m$ which can result in a homogeneous
polynomial (\ref{Fm0}) is $\tilde{F}_0$.

\section{Conclusion}

We have discussed mirror symmetry of supermanifolds constructed
as fermionic extensions of toric varieties. This has been achieved by
studying fermionic extensions of linear sigma A-models and their
T-dual super-LG B-models. The present work primarily concerns
the quadratic case where $n=f$ (i.e., equal numbers of $U(1)$ factors
and fermionic fields in the A-model), and focus has been on the
bosonic structure arising after integrating out the fields in the B-model
obtained by dualizing the fermionic fields in the A-model.
We have found that quasi-homogeneity of the resulting toric data
of the B-model is related to the super-CY conditions of the A-model
supermanifold. Furthermore, the degree of the associated B-model
superpotential is given in terms of the determinant of the A-model fermion
charge matrix. Several special cases have been used as illustrations
of our general results.

Natural extensions of the present work include the non-quadratic case
where $n\neq f$. It is also of interest to understand the different
patches of the bosonic B-model structure that would result after
integrating out different bosonic fields than the ones introduced
by the dualization of the fermionic fields in the A-model.
One should also try to extract geometric information on the
full supermanifold in the B-model, and not just the bosonic
structure of it obtained after integrating out the fermionic fields.
We hope to discuss all of these interesting problems in the
future \cite{Work}.
\\[.4cm]
\noindent{\em Acknowledgments}
\\[.2cm]
Belhaj thanks R. Ahl Laamara and P. Resco for discussions, while
Rasmussen thanks C. Cummins for discussions. Drissi and Saidi thank
the program Protars III, CNRT D12/25, Rabat.

\end{document}